\documentclass[12pt,aip,jap,preprint,superscriptaddress,floatfix,showkeys]{revtex4-1}
\usepackage{graphicx}
\usepackage{float}
\usepackage{times}
\usepackage{amsfonts,amsmath}
\usepackage{enumitem}

\newcommand{\Borja}{Borja Esteve-Altava}
\newcommand{\Valles}{Toni Vall\`es-Catal\`a}
\newcommand{\Diego}{Diego Rasskin-Gutman}
\newcommand{\Guimera}{Roger Guimer\`a}
\newcommand{\Sales}{Marta Sales-Pardo}
\newcommand{\URV}{Departament d'Enginyeria Qu\'{\i}mica, Universitat Rovira i 
Virgili, 43007 Tarragona, Catalonia}
\newcommand{\ICREA}{Instituci\'o Catalana de Recerca i Estudis Avan\c{c}ats 
(ICREA), Barcelona 08010, Catalonia}
\newcommand{\HUCM}{Department of Anatomy, Howard University College of 
Medicine, Washington, DC, United States of America} 
\newcommand{\RVC}{Structure \& Motion Laboratory, Department of Comparative 
Biomedical Sciences, Royal Veterinary College, London, United Kingdom}
\newcommand{\UV}{Theoretical Biology Research Group, Cavanilles Institute of 
Biodiversity and Evolutionary Biology, University of Valencia, Valencia, Spain}

\begin{document}
\title{Bone fusion in normal and pathological development is constrained by the network architecture of the human skull}
\author{\Borja}
\thanks{Co-first author}
\affiliation{\RVC}
\affiliation{\HUCM}
\affiliation{\UV}
\author{\Valles}
\thanks{Co-first author}
\affiliation{\URV}
\author{\Guimera}
\affiliation{\ICREA}
\affiliation{\URV}
\author{\Sales}
\thanks{Corresponding author: marta.sales@urv.cat}
\affiliation{\URV}
\author{\Diego}
\affiliation{\UV}

\begin{abstract}
\textbf{Abstract -} The premature fusion of cranial bones, craniosynostosis, affects the correct development of the skull producing morphological malformations in newborns. To assess the susceptibility of each craniofacial articulation to close prematurely, we used a network model of the skull to quantify the link reliability (an index based on stochastic block modeling and Bayesian inference) of each articulation. We show that, of the 93 human skull articulations at birth, the few articulations that are associated with nonsyndromic craniosynostosis conditions have statistically significant lower reliability scores than the others. In a similar way, articulations that close during the normal postnatal development of the skull have also lower reliability scores than those articulations that persist through adult live. These results indicate a relationship between the architecture of the skull network and the specific articulations that close during normal development and in pathological conditions. Our findings suggest that the topological arrangement of skull bones might act as an epigenetic factor, predisposing some articulations to closure, both in normal and pathological development, and also affecting the long-term evolution of the skull.
\end{abstract}


\maketitle

\section*{Introduction}
Craniofacial articulations are primary sites of bone growth and remodeling; adequate formation and maintenance of these articulations is therefore important for a healthy development of the head and brain. The timely closure of bone articulations is a normal process that takes place during skull development. Craniosynostosis is a pathological condition in which one or more articulations between cranial bones (frontal, parietal, temporal, and occipital) close prematurely, leading to the fusion of these bones. Craniosynostosis has an estimated prevalence of about 5 in 10,000 live births \citep{di2009evolution}. The premature fusion of bones, if not treated surgically, can cause head malformations due to compensatory growth of other joints \citep{delashaw1989cranial}, sometimes provoking severe brain damage due to an increase of intracranial pressure \citep{inagaki2007intracranial}. Craniosynostosis can occur in isolation, as nonsyndromic craniosynostosis \citep{garza2012nonsyndromic,watkins2014classification}, or as part of a variety of congenital disorders, such as Apert and Crouzon syndromes \citep{rice2008clinical}. In general, it is not well understood what factors predispose some articulations but not others to close pathologically or in normal development.

Genetic and epigenetic factors participate in the formation and maintenance of craniofacial articulations through life. The number of genes identified to be carrying mutations associated with craniosynostosis has grown in the last two decades \citep{twigg2015genetic}. For example, more than 60 genes are now known to carry mutations associated with craniosynostosis \citep{twigg2015genetic}: some of them show specificity for a suture in the context of a syndrome (e.g., \textit{ASXL1} and metopic suture in the Bohring-Opitz syndrome), others predispose to more than one type of craniosynostosis (e.g., \textit{FGFR2} in coronal, sagittal, and multisuture synostoses), while most of them are not specifically associated with suture development, but to osteogenesis in general (e.g., \textit{ALX4}, \textit{EFNA4}, and \textit{TGFBR2}). Epigenetic factors include, among many others, bio-mechanical stress, hypoxia, and use of drugs during pregnancy\citep{oppenheimer2009force,percival2011epigenetics,watkins2014classification}. Thus, epigenetic factors are even less specific than genetic ones; for example, maternal smoking has been associated to a predisposition for various craniosynostoses \citep{carmichael2008craniosynostosis,shi2008review}. 

\begin{figure}[!ht]
  \centering
  \includegraphics[scale=0.65]{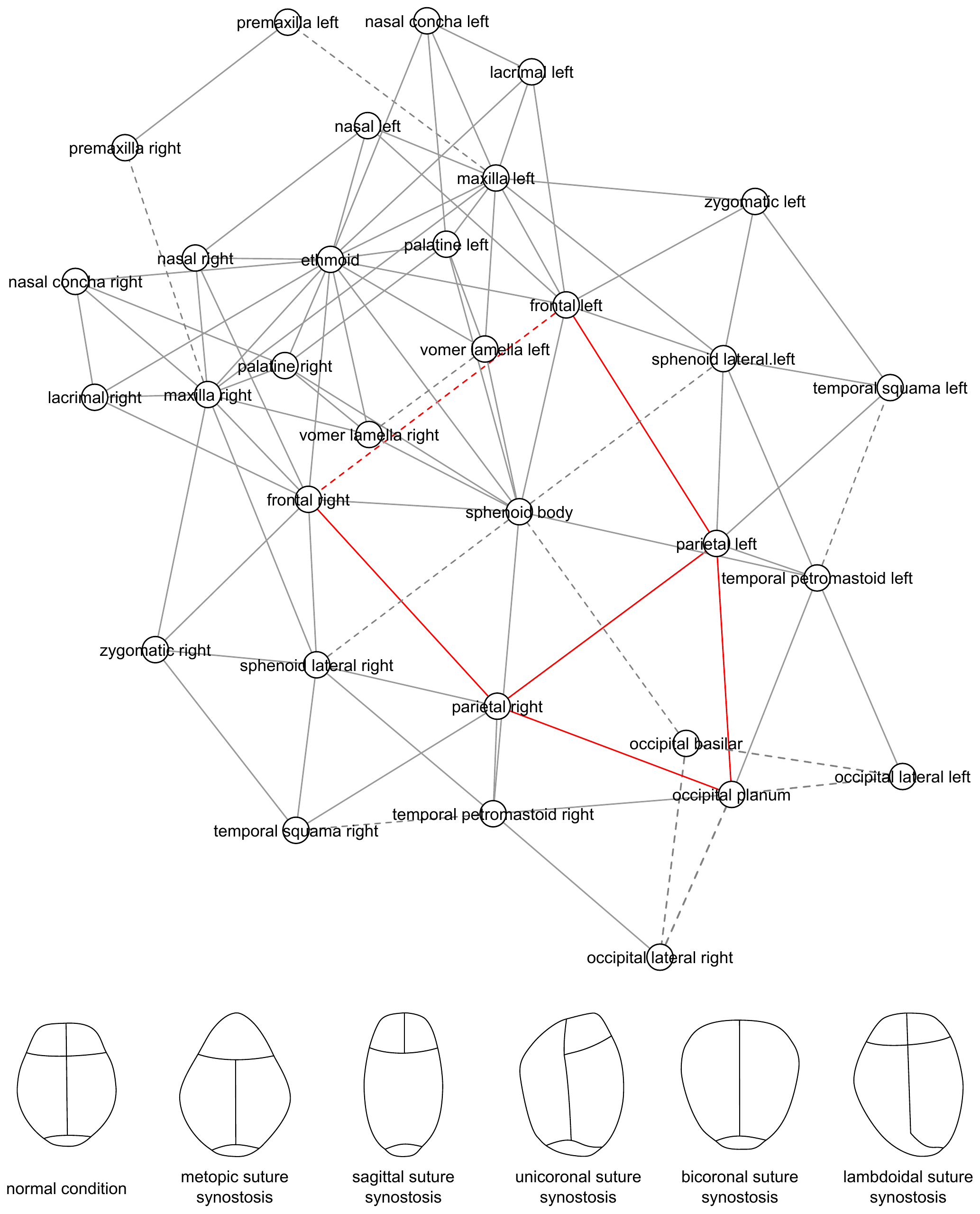}
  \caption{\label{fig1}The arrangement of bones in the human skull at birth modeled as a network (\textit{top}). Nodes represent bones and links represent articulations among bones (cartilaginous and fibrous joints). Red links are articulations associated with craniosynostosis conditions; dashed links are articulations that close during the normal development of the skull. Note that the metopic suture between the left and right frontal bones closes in both pathological and normal development. Drawings illustrate the shape of the head in some of the conditions studied (\textit{bottom}).}
\end{figure}

We addressed the articulations susceptibility to close from a theoretical standpoint, by modeling the skull as a network in which nodes and links formalize bones and their articulations at birth (Fig.~\ref{fig1}). Anatomical network models have been used before, for example, to identify developmental constraints in skull evolution \citep{esteve2013structural,esteve2014random}, analyze the evolution of tetrapod disparity in morphospace across phylogeny \citep{esteve2014theoretical}, and model the growth of human skull bones \citep{esteve2014beyond}. A recent comparison of network models of craniosynostosis conditions showed that, despite the associated abnormal shape variation, skulls with different types of craniosynostosis share a same general pattern of network modules \citep{esteve2015evo}.  

Using the reliability formalism developed for network models \citep{guimera2009missing} we infer the susceptibility of craniofacial articulations to close prematurely. A common feature of the topology of complex networks such as the skull is that one can identify groups of nodes (bones) that have well-defined patterns of connections (i.e., articulations) with other groups of nodes \citep{guimera2009missing}. Such realization allows one to identify links that are topologically unexpected. If the architecture of the skull is driving the closure of articulations, we surmise that there is a relationship between the susceptibility of a pair of bones to fuse and the topological 'unexpectedness' of their articulation. To quantify such susceptibility, we use the \textit{link reliability} score, that is the probability that a connection exists in the network given the observed (neonatal) topology of the skull \citep{guimera2009missing}. A low score means that the presence of this articulation is rare, that is, not commonly expected in the given arrangement of bones (see \textit{Methods} for details on how this is estimated). Importantly, the link reliability formalism  has been used in other complex systems to accurately predicting missing and spurious interactions in social, neural, and molecular networks \citep{guimera2009missing}, to predict harmful interactions between pairs of drugs \citep{guimera2013network}, and to predict the appearance of conflicts in teams \citep{rovira2013predicting}. Here we use the reliability formalism to investigate whether the topological arrangement of bones predicts which articulations are more susceptible to close in development; in other words, we want to assess if the architecture of the skull acts as an agent that constrains the fusion of bones.

\newpage

\newpage

\section*{Methods}
\subsection*{A network model of the skull}

We built a network model of the human skull at birth based on anatomical descriptions \citep{gray1918anatomy} and information of ossification timing and fusion events \citep{sperber2010craniofacial}. The nodes and links of the network model formalize the bones and articulations of the skull, respectively (Fig.~\ref{fig1}). For simplicity, we use \textit{bone} in a broad sense to refer both to bonny elements (e.g., a parietal bone) and well-formed cartilaginous templates of the future bones (e.g., the ethmoidal bone). Likewise, we use \textit{articulation} to refer to cartilaginous (synchondroses) as well as fibrous joints (craniofacial sutures). We are aware that each type of skeletal element and articulation has different biological properties, which might be hardly comparable in some contexts. However, our theoretical analysis focuses at a higher level of abstraction, that of topology (i.e., the arrangement of constituent parts), aiming to extract relevant information from the sole topological structure of the skull. Thus, specific properties of nodes (e.g., cellular origins, ossification mechanisms) and of articulations (e.g., contact areas, tensile properties) have not been included in the present model \citep[see][for a review of examples of how anatomical network analysis abstractions have successfully been applied in different anatomical contexts]{esteve2015anatomical,esteve2015anatomical2,rasskin2014connecting}.

\subsection*{Topological organization of the neonatal skull}
The topological organization of the skull varies during pre- and postnatal development. We have chosen to work with the skull configuration at birth because it allows a broader comparison between closed and persistent articulations, both in normal and pathological conditions. What follows is a summary of the bones present at birth that we used to build the neonatal skull network model \citep[see][for details]{gray1918anatomy,sperber2010craniofacial}.

The occipital bone at birth consists of four units: a ventral basilar part, a more dorsal occipital plate, and two lateral parts. Around the fourth year the occipital plate and the lateral parts fuse into one unit. Around the sixth year the basilar part is also fused together. During adulthood (about 18-25 years) the occipital bone and the sphenoid bone fuse into a single unit. The frontal bone at birth consists of two halves separated by the metopic suture. Around the eighth year the metopic suture obliterates and the two halves of the frontal fuse into one single bone (although in some individuals the suture endures and left and right frontals are present through life). The premature fusion of the metopic suture is one of the craniosynostosis conditions included in the present study (see Fig.~\ref{fig1}). Each temporal bone at birth consists of two parts: the petromastoid and the squama (to which the tympanic ring has united shortly before birth). Around the first year the petromastoid and squama fuse into a single unit. The temporal bone has a tight relationship with two small structures: the ear ossicles (maellus, incus, and stapes) and the styloid process (tympanohyal part and stylohyal part). The former structures develop partially embedded within the temporal bone, while the latter structures fuse with it during the first years of development. For simplicity, we have decided not to consider these structures as separate nodes in the network model; instead, we include them within the temporal bone in order to focus on the main skeletal units of the skull. The sphenoid bone at birth consists of three parts: a central body (including the small wings) and two lateral parts or alisphenoids (comprising the great wings and the pterygoid processes). Around the first year the sphenoid body and the alisphenoids fuse together. As we already mentioned, the sphenoid and the occipital fuse into a single unit during adulthood. The ethmoid bone is still a cartilaginous template at birth, which will later ossify endochondrally to form the ehtmoid bone. The maxilla and premaxilla (one of each per side) at birth are still separated by a suture that can persist until well into adulthood. Each zygomatic bone consists of one single skeletal structure at birth, although sometimes can be divided horizontally in an upper and a lower part (a similar division is reported by Gray to occur in \textit{"quadrumana"} [sic]), which would indicate that this phenotype might be an atavism). The vomer at birth consist of two lamellae, which fuse together at puberty (although it might be still traces of their paired laminar origin). Finally, the lacrimals, nasals, inferior nasal conchae, palatines, and parietals are well-formed skeletal units at birth (although the parietal and palatines still will continue growing some time after birth). Gray reports that, at times, the parietal bone can be divided by a longitudinal suture in an upper and a lower part (as this is a deviation of the more common pattern found in humans, we did not includ this phenotype in our network model).

\subsection*{Estimation of link type probability using stochastic block models}

Stochastic block models are good models to describe the patterns of connections in complex networks. In such models, nodes are assigned to groups and the probability of a link existing between two pairs of nodes is given by a matrix that specifies the connectivity rate between nodes belonging to pairs of blocks. For a given network, good stochastic block models are those that group nodes that have a similar pattern of connections; for instance, in our case we could group together nodes $vomer$ and $palatine$ since both tend to connect to similar nodes ($sphenoid$, $ethmoid$, $maxilla$) along with a disconnection to similar nodes (e.g., $parietal$, $zygomatic$, $frontal$). Within this description, links between pairs of nodes that belong to groups 
that are densely interconnected are more likely than those links between pairs of nodes belonging to groups that are sparsely connected. For instance, in the previous example an articulation existing between $palatine$ and $maxilla$ is much more likely than a suture between $palatine$ and $parietal$.

To mathematically formalize this intuition, we compute the reliability score, that is the probability that a link exists given the network of connections we observe (the newborn skull in our case) using stochastic block models as the basis for our inference algorithm. In practice, our algorithm samples the space of partitions of nodes into groups taking into account how good a given partition manages to classify nodes with similar patterns of connections into the same group. For each of these partitions, each link between a pair of nodes $(i,j)$ has a specific probability. The reliability score of link $N_{ij}$ is then a weighted average of the probabilities of that link for each sampled partition. Mathematically, we formalize the previous arguments in a Bayesian framework as follows. Given a family of models $\mathcal{M}$, the probability that $N_{ij}=1$ 
given the observed network $N^O$ (that is the matrix of connections) is \citep{guimera2009missing}
\begin{equation}
  p(N_{ij}=1|N^O)=\int_\mathcal{M} dM \, p(N_{ij}=1|M) \, p(M|N^O) \,,
\end{equation}
where the integral is over all the models $M$ in ensemble  $\mathcal{M}$.
We can rewrite this equation using Bayes theorem and obtain 
\citep{guimera2009missing,guimera2012predicting}
\begin{equation}
  p(N_{ij}=1|N^O)=\frac{\int_\mathcal{M} dM \, p(N_{ij}=1|M) \, p(N^O|M) \, 
p(M)}{\int_\mathcal{M} dM \, p(N^O|M) \, p(M)} \;.
  \label{e-bayes}
\end{equation}
Here, $p(N^O|M)$ is the probability of the observed interactions given model $M$ and $p(M)$ is the {\it a priori\/} probability of a model, which we assume to be model-independent $p(M)={\rm const}$. In our approach, we assume that the family of stochastic block models is a good ensemble to describe the connectivity in a complex network (in our case that of the human skull). Therefore, each model $M=(P, Q)$ is completely determined by a partition $P$ of bones into groups and the group-to-group 
interaction probability matrix $Q$. For a given partition $P$, the matrix element $Q(\alpha, \beta)$ is the probability of an articulation joining a bone in group $\alpha$ with a bone in group $\beta$. Thus, if $i$ belongs to group $\sigma_i$ and $j$ to group $\sigma_j$ we have that \citep{guimera2012predicting}
\begin{equation}
  p(N_{ij}=1|M)=Q(\sigma_i, \sigma_j) \;;
  \label{e-p1}
\end{equation}
and
\begin{equation}
  p(N^O|M)=\prod_{\alpha \le \beta} Q(\alpha, \beta)^{n^1(\alpha, 
\beta)}(1-Q(\alpha, \beta)^{n^0(\alpha, \beta)}) \;,
  \label{e-p2}
\end{equation}
where $n^1(\alpha,\beta)$ is the number of articulations between bones in groups $\alpha$ and $\beta$ and $n^0(\alpha,\beta)$ is the number of disconnections between bones in groups $\alpha$ and $\beta$.

The integral over all models in $\mathcal{M}$ can be separated into a sum over all possible partitions of the bones into groups, and an integral over all possible values of each $Q(\alpha,\beta)$. Using this together with Eqs.~(\ref{e-bayes}), (\ref{e-p1}) and (\ref{e-p2}), and under the assumption of no prior knowledge about the models ($p(M)={\rm const.}$), we have
\begin{eqnarray}
&&  p(N_{ij}=1|N^O) = \\
&&  \frac{1}{Z} \sum_{P} \int_0^1 dQ \hspace{1mm} Q(\sigma_i, \sigma_j) 
\prod_{\alpha \le \beta} Q(\alpha, \beta)^{n^1 
(\alpha,\beta)}(1-Q(\alpha,\beta)^{n^0(\alpha,\beta)})\;, \nonumber
 \label{e-derivation1}
\end{eqnarray}
where the integral is over all $Q(\alpha, \beta)$ and $Z$ is the normalizing constant (or partition function). Using these expressions in Eq.~(\ref{e-derivation1}), one obtains
\begin{equation}
  p(N_{ij}=1|N^O)=\frac{1}{Z} \sum_P \left( 
\frac{n^1(\sigma_i,\sigma_j)+1}{n(\sigma_i,\sigma_j)+2} \right) \exp (-H(P)) \;,
  \label{e-p}
\end{equation}
where the sum is over all partitions of bones into groups, $n(\sigma_i,\sigma_j)= 
n^1(\sigma_i,\sigma_j) + n^0(\sigma_i,\sigma_j)$ is the total number of 
possible sutures between groups $\sigma_i$ and $\sigma_j$, and $H(P)$ is a 
function that depends on the partition only
\begin{equation}
  H(P)= \sum_{\alpha \le \beta} \left[ \ln (n(\alpha,\beta) + 1) + \ln 
\binom{n(\alpha,\beta)}{n^1(\alpha,\beta)} \right] \,,
  \label{e-H}
\end{equation}
This sum can be estimated using the Metropolis algorithm \citep{metropolis1953equation,guimera2009missing} as detailed next.

\subsection*{Implementation details}
The sum in Eq. (\ref{e-p}) cannot be computed exactly because the number of 
possible partitions is combinatorially large, but can be estimated using the 
Metropolis algorithm \citep{metropolis1953equation,guimera2009missing}. This 
amounts to generating a sequence of partitions in the following way. From the 
current partition $P^0$, select a random bone and move it to a random new group 
giving a new partition $P^1$. If $H(P^1)<H(P^0)$, always accept the move; 
otherwise, accept the move only with probability $P=e^{H(P^0)-H(P^1)}$.
By doing this, one gets a sequence of partitions $\{P^i\}$ such that one can approximate the integral in Eq. \ref{e-p} as 
\citep{metropolis1953equation}
\begin{equation}
p(N_{ij}=1|N^O) \approx \frac{1}{S} \sum_{P \in \{P^i\}} 
\frac{n^1(\sigma_i,\sigma_j)+1}{n(\sigma_i,\sigma_j)+2}\;,
  \label{e-pmcmc}
\end{equation}
where $S$ is the number of sampled partitions in $\{P^i\}$.

In practice, it is useful to ``thin'' the sample $\{P^i\}$, that is, to 
consider only a small fraction of evenly spaced partitions so as to avoid the 
computational cost of sampling very similar partitions which provide very 
little additional information. Moreover, one needs to make sure that sampling 
starts only when the sampler is ``thermalized'', that is, when sampled 
partitions are drawn from the desired probability distribution (which in our 
case is given by $e^{-H(P)} / Z$). Our implementation automatically determines 
a reasonable thinning of the sample, and only starts sampling when certain 
thermalization conditions are met. Therefore, the whole process is completely 
unsupervised. The source code of our implementation of the algorithm is 
publicly available from 
http://http://seeslab.info/downloads/network-c-libraries-rgraph/ and 
http://github.com/seeslab/rgraph.

\subsection*{Statistical analysis}

We performed independent Mann-Whitney U tests for the following comparisons: (1) articulations affected by nonsyndromic craniosynostosis \textit{vs.} articulations unaffected; and (2) articulations normally closed in development \textit{vs.} articulations that persist in the adult; and (3) articulations that close in craniosynostosis \textit{vs.} articulations that close during normal development. The effect size of the difference of means between groups in standard deviations was estimated using the Cohen's \textit{d}. The statistical analysis was performed using JASP version 0.7.5.6.

We tested the null hypothesis of equal distribution between groups against the corresponding alternative hypotheses that:
\begin{enumerate}[topsep=0pt, partopsep=0pt]
\itemsep0em 
 \item articulations affected by craniosynostosis have lower reliability scores than articulations unaffected (one-sided test); 
 \item articulations that close during normal development have lower reliability than those that persist in the adult (one-sided test);
 \item articulations affected by craniosynostosis have different reliability scores than those that close during normal development (two-sided test).
\end{enumerate}

\section*{Results}

The human skull at birth comprises 32 bones and 93 articulations, of which only a small fraction are associated with nonsyndromic craniosynostosis conditions. We investigated the relationship between the link reliability score and the susceptibility of an articulation to close during normal development or due to craniosynostosis.

\begin{figure}[!ht]
  \centering
  \includegraphics[scale=0.5]{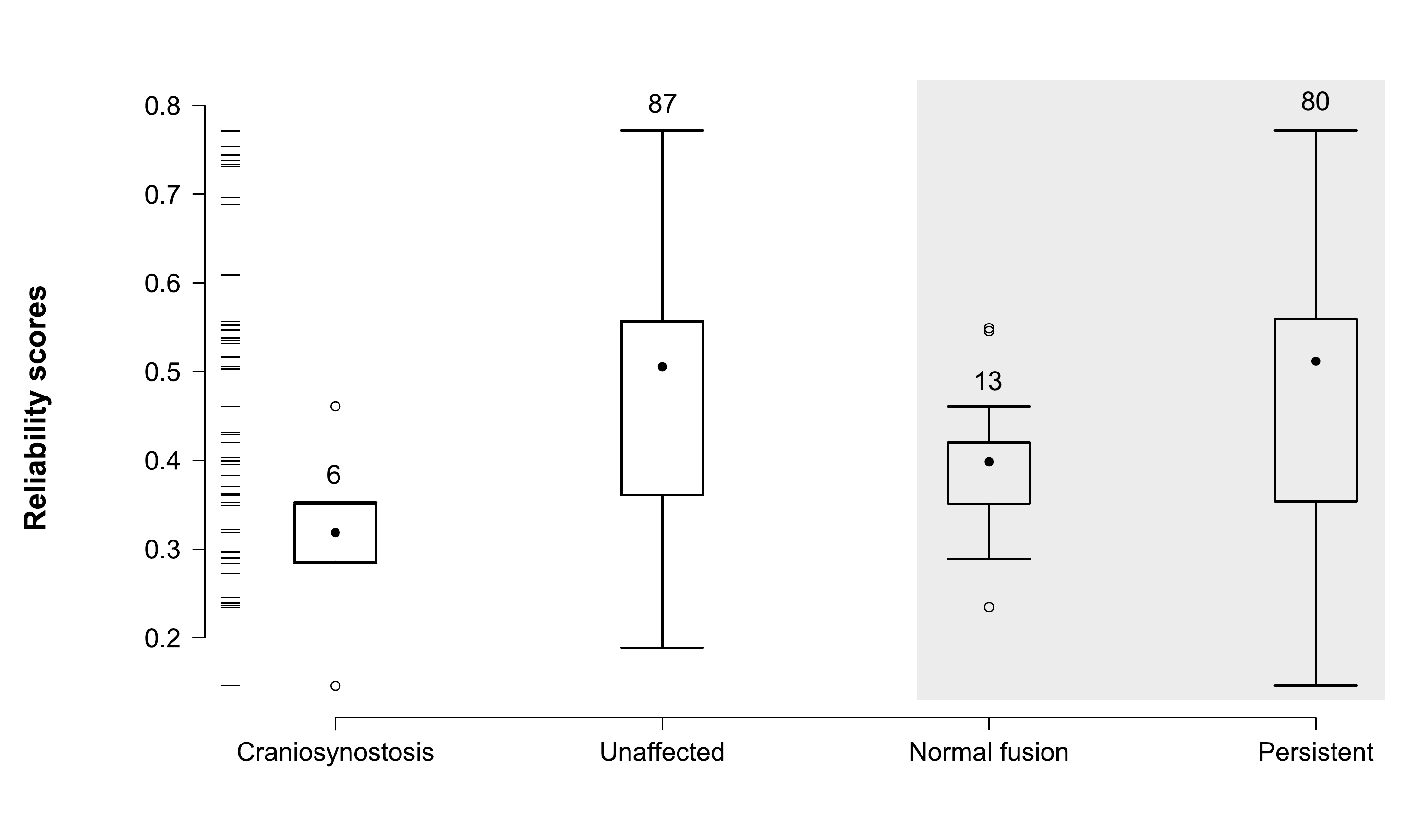}
  \caption{\label{fig2}Box plot comparing link reliability scores. Articulations associated with craniosynostosis have lower reliability than those that are not associated (\textit{left}, white panel). Articulations that close during normal development also have lower reliability than those that will persist in the adult live (\textit{right}, gray panel).}
\end{figure}

First, we compared the reliability score of those articulations that close during the normal development of the skull to those that persist in the adult. We find that sutures that normally close have significantly slightly lower reliability scores than those that do not 
(Mann-Whitney-Wilcoxon: one sided, W=368, p-value = 0.047; Cohen's \textit{d} = -0.52) (Fig.~\ref{fig2}); 
which is in agreement with our hypothesis that during normal development there is a tendency to close articulations that are topologically rare in the newborn skull.

Next, we compared the reliability score of articulations that close prematurely in craniosynostosis to that of those articulations unaffected by this pathological condition (Fig.~\ref{fig2}). We found that articulations associated with craniosynostosis have significantly lower reliability scores than unaffected articulations 
(Mann-Whitney-Wilcoxon: one-sided, W = 98, p-value = 0.006; Cohen's \textit{d} = -1.066) (Fig.~\ref{fig2}); 
which shows that articulations associated to craniosynostosis are also unexpected from a topological point of view.

Interestingly, we find also that the reliability scores of articulations that close in craniosynostosis conditions are not statistically different than those that close during normal development 
(Mann-Whitney-Wilcoxon: one sided, W=15.5, p-value = 0.087; Cohen's \textit{d} = -0.964). 
This finding suggests that while skull architecture is an important factor in the loss of sutures during both pathological and normal development, there are non-topological factors that discriminate between normal and pathological loss of sutures. However, this result must be interpreted with caution due to the small sample size of both groups ($N=6$ and $N=11$, respectively); notice that the Cohen's \textit{d} is in fact indicating a difference of means of a similar magnitude to that observed in the previous comparison (see also Fig.~\ref{fig2}). Further details of the statistical analysis and score values are available in the Supplementary Information.

\section*{Discussion}

Our results suggest that the whole arrangement of craniofacial articulations of the skull might act itself as an epigenetic factor, making some articulations to be more susceptible to closure than others. That some regions of the skull act epigenetically (e.g., via bio-mechanical signaling) to predispose bones to a premature fusion was already proposed by Moss in the context of the functional matrix hypothesis \citep{moss1975functional}. Here we show that the most susceptible articulations to close prematurely (i.e., those with low reliability scores) are precisely the ones associated with craniosynostosis. Thus, we propose that the very arrangement of bones in the skull predisposes epigenetically some articulations as targets of pathological conditions. We are not yet in a position to offer a mechanistic explanation for the relationship reported here, which we believe may be related to the same developmental mechanism that regulate compensatory growth of bones after premature synostoses \citep{delashaw1989cranial,morriss2005growth,lieberman2011evolution}. However, our results also suggest that such mechanisms might not be different between normal development and pathological conditions, since articulations that close during normal development also show low reliability scores compared to those articulations that persist in the adult skull.

If, as our results suggest, the system of articulations of skull bones is able to self-regulate epigenetically the formation and maintenance of individual bone articulations, this might have consequences also at an evolutionary scale. In craniosynostosis conditions, the number of bones is reduced due to the early fusion of bones, much in the same way as the net reduction in the number of bones during vertebrate evolution \citep{gregory1935williston,sidor2001simplification,esteve2013structural}; as a 
consequence, it has been postulated that craniosynostosis could be used as an informative model for skull evolution \citep{richtsmeier2006phenotypic}. Our results suggest that this is not a mere analogy, but that similar epigenetic 
processes might act in regulating (or constraining) the configuration of bone arrangements in the skull, both in development and in evolution.

Pathological conditions of the human skull such as craniosynostosis are a medical and social problem that needs special attention from the research community. In addition, they represent medical examples of more general developmental and evolutionary processes found in all tetrapods. Both aspects, the medical and the biological, need and can be integrated in order to reach a better understanding that could lead to improve treatments as well as to further our knowledge about fundamental evolutionary questions.

\section*{Acknowledgments}
This project has received funding from the European Union’s Horizon 2020 
research and innovation programme under a Marie Sklodowska-Curie grant (654155) 
to BE-A (654155), from the Ministerio de Economia y Competitividad de Espa\~na 
(BFU2015-70927-R) to DR-G and BE-A, from the Ministerio de Economia y Competitividad de Espa\~na (FIS2013-47532-C3-P1) to RG \& MS-P, from FP7 FET-Proactive 317532 to RG \& MS-P and from the James S McDonnell Foundation to RG \& MS-P.

\section*{Author's contribution}
All authors designed the study. BE-A made the network model of the skull. RG, MS-P, and TV-C analyzed the network model and calculated reliability scores. All authors discussed the results and wrote the manuscript.

\bibliography{csreliabilityreferences}
\end{document}